\documentclass[showpacs,prl,floatfix,twocolumn,amsmath,a4]{revtex4-1}
\usepackage{graphicx}

\begin{document}

\title{Correlation-driven electronic multiferroicity in (TMTTF)$_2$-$X$ organic crystals}
  
\author{Gianluca Giovannetti$^{1}$, Reza Nourafkan$^{2}$, Gabriel Kotliar$^{2}$, and Massimo Capone$^{1}$}
\affiliation{$^1$ CNR-IOM-Democritos National Simulation Centre and International School
for Advanced Studies (SISSA), Via Bonomea 265, I-34136, Trieste, Italy }
\affiliation{$^2$ Department of Physics and Astronomy, Rutgers University, Piscataway, NJ 08854-8019, USA }
\date{\today}

\begin{abstract}
Using a combination of density functional theory and dynamical mean field theory we show that electric polarization and magnetism are strongly intertwined in (TMTTF)$_2$-$X$ (X$=$PF$_6$, As$F_6$, and SbF$_6$) organic crystals and they originate from short-range Coulomb interactions. 
Electronic correlations induce a charge-ordered state which, combined with the molecular dimerization, gives rise to a finite electronic polarization and to a ferroelectric state. We predict that the value of the electronic polarization is enhanced by the onset of antiferromagnetism showing a sizable magnetoelectric leading to a multiferroic behavior of (TMTTF)$_2$-$X$ compounds.
\end{abstract}


\maketitle 

In the last decade the study of ferroelectric and multiferroic materials is emerging as a novel frontier in condensed matter physics for their wide range of potential applications\cite{Cheong}. 
Ferroelectricity appears when ionic or molecular distortions break the inversion symmetry and the coherent orientation of dipole moments creates a net polarization. Despite the traditional mechanisms of magnetism and ferroelectricity are typically  incompatible\cite{Hill}, the simultaneous ordering of electrical and magnetic degrees of freedom is possible and defines multiferroic materials.  We can classify multiferroics according to the microscopic mechanism that determines their properties\cite{Khomskii}, and to the relation between the two orderings. In particular, we have materials in which ferroelectricity and magnetism have different origin,  but also systems in which magnetism controls ferroelectricity or even causes it. These latter multiferroics are promising candidate for observing sizable room-temperature magneto-electric responses, which can pave the way to the development of magnetic devices which can be switched by applying an external voltage\cite{Vaz}. A further promising direction is the development of {\it{electronic ferroelectrics}}, in which the polarization has a predominant electronic contribution\cite{Kobayashi, Portengen96, Batista02}.

An electronic mechanism compatible with both ferroelectric and magnetic orderings is based on charge ordering\cite{KhomskiiBrink}. In charge-ordered (CO) systems, the coexistence of inequivalent bonds and inequivalent sites with different carrier density leads to a ferroelectric state, which can be multiferroic if the CO state also supports magnetic ordering. The realization of a similar mechanism in the Fabre charge-transfer organic salts (TMTTF)$_2$-$X$ has been hinted\cite{KhomskiiBrink,Giovannetti}, but the role of strong electron-electron correlation in the complex interplay between electric polarization and magnetism has not been investigated and elucidated so far. As in other molecular solids, the screened Coulomb interaction is expected to play a major role because of the narrow bands arising from the overlap between molecular orbitals. 

Here we use a combination of density functional theory (DFT) \cite{DFT} and dynamical mean-field theory (DMFT) \cite{DMFT} to study the correlation-driven emergence of a ferroelectric state in (TMTTF)$_2$-$X$ and its interplay with the magnetic order evaluating the electronic contribution to the polarization with a recently introduced method based on the interacting Green's functions\cite{RezaNourafkanPdmft}. We find that short-range correlations give rise to a characteristic coupling between magnetism and polarization and consequently to a multiferroic state.

The family of (TMTTF)$_2$-$X$ organic salts displays indeed diverse electronic properties that can be controlled by substituting the counterion X or by applying pressure and their phase diagram include various electronic phases such as charge ordering, spin density wave and antiferromagnetism (AFM)\cite{Brazovskii,Jerome}. 
In particular, at least three members of the family, $X=$PF$_6$, As$F_6$, and SbF$_6$ develop a low-temperature CO state with a 4K$_F$ modulation, with critical temperature 67, 102, 157 K respectively, which in turn coincides with the onset of the ferroelectric order \cite{Monceau,Chow}. At very low temperatures T $\lesssim 17K$, a spin-Peierls state establishes for $X=$PF$_6$ and As$F_6$ while for $X=$SbF$_6$ the CO state in coexist with an AFM phase  at T $\lesssim 8K$\cite{Matsunaga}. The magnetic response is not trivial also at higher temperatures, where the electron-spin resonance spectra can be described by a spin-$1/2$ antiferromagnetic Heisenberg chain for all the three compounds we consider\cite{Dumm}.  The existence of localized spins above the ordering temperature is a signature of Mott localization, whose interplay with CO and magnetic ordering will be shown to be the key to the electronic multiferroic behavior. 

The crystal structure is characterized by TMTTF molecules stacked along the $a$ axis, separated by the X counterion, and weakly interacting along the $b$ axis, leading to a quasi-onedimensional bandstructure.  In the following we will use the electronic structure of (TMTTF)$_2$-PF$_6$, for which accurate structural data are available, as a baseline for the analysis of the whole family. It is important to notice that in the CO state the crystals preserve the centrosymmetric structure\cite{Jacko} despite the molecular dimerization, suggesting a small ionic contribution to the polarization. 

We performed DFT calculations in the Perdew-Burke-Ernzerhof (PBE)\cite{PBE} scheme using  Quantum Espresso\cite{QE}. A two-dimensional tight-binding representation of the bandstructure is built using Wannier90\cite{W90}. This mapping from DFT electronic structure to the localized Wannier molecular orbitals of X$=$PF$_6$ is representative of all the members of the charge-transfer (TMTTF)$_2$-$X$ as indeed changes in the hopping parameters along the series are rather small.

The low-energy electronic structure of all the members of the family (TMTTF)$_2$-$X$ is characterized by two bands arising from TMTTF HOMO orbitals\cite{Jacko}. The conduction bands arise from the highest occupied molecular orbital s of the two inequivalent TMTTF molecules and they have a width $W \sim 1.0$ eV \cite{Giovannetti}. The ratio 2:1 between cations (TMTTF molecules) and anions (PF$_6$ group) in the unit cell leads to a commensurate band filling of $3/4$ which leads to a metallic state within PBE\cite{Jacko}.  We then include the interactions as described by the Hamiltonian
\begin{equation}
H = \sum_{\langle i,j\rangle \sigma} t_{ij}( {\xi})  c^{\dagger}_{i \sigma}  c^{~}_{j \sigma} 
+ U \sum_{i} n_{i \uparrow} n_{i \downarrow}
+ V \sum_{\langle i,j\rangle} n_{i} n_{j},
\label{Ham}
\end{equation}
where $c_{i \sigma}$ and $c^{\dagger}_{i \sigma}$ are annihilation and creation operators for electrons at site $i$ with spin $\sigma$, $t_{ij}$ denotes the hopping parameters. The hopping is two-dimensional, while the non-local repulsion is restricted to sites along the same chain. 
For the sake of clarity, we parameterize the distortion in terms of the deviation of the hoppings along the one-dimensional chains with respect to their undistorted value of $t_0 =$ 0.21 eV. When we approach the actual dimerized structure, the two inequivalent hoppings become $t_{\pm} = t_0 \pm \xi$, with $\xi =$ 0.01 eV in the fully distorted structure. $U$ is the on-site Hubbard repulsion and $V$ is an inter-site repulsion between TMTTF sites along the stacking direction. We mention here that the inclusion of interchain hopping parameters does not affect the interplay between CO and magnetism, while it is important to determine the actual two-dimensional magnetic and charge pattern\cite{Yoshimi}.

We perform DMFT calculations for different lattice distortions (or $\xi$) and we compute the Green's function and the electronic polarization according to Ref. \cite{RezaNourafkanPdmft}. The interaction $V$ is  treated in the Hartree approximation and the DMFT equations are solved with a zero-temperature exact-diagonalization  impurity solver \cite{Caffarel,Capone} using an Arnoldi algorithm. We allow for charge- and spin-ordering introducing four independent sites. For each of them we define an impurity model and we compute a local dynamical self-energy.  
In order to identify the relative role of the two interaction terms, we vary $U$ and $V$ respectively in a range around previous estimates of $U$=2.2 eV\cite{Giovannetti} and $V$=0.4 eV\cite{Mila}.

\begin{figure}
\includegraphics[width=1.0\columnwidth,angle=0]{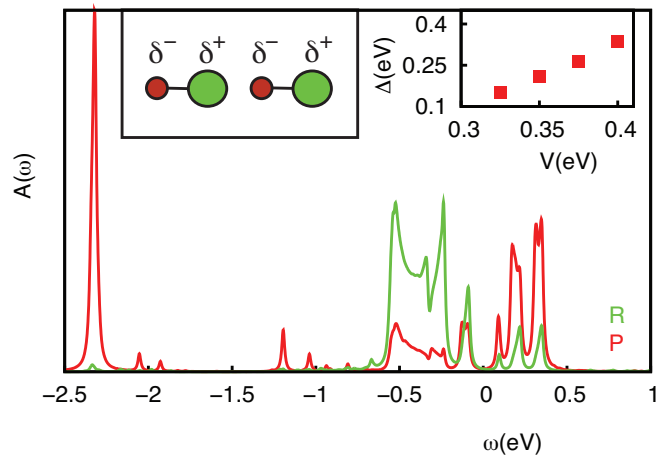}
\caption{(Color online) Spectral density of states calculated in the paramagnetic state for U$=$2.2 and V$=$0.325 within DFT+DMFT scheme. Insets: (Left) Charge arrangement of poor (P) and rich (R) sites in the CO paramagnetic state along the stacking direction $a$ at ${\xi}$=0.01 ; (Right) Charge gap ${\Delta}$ as function of V.}
\label{fig1}
\end{figure}

We start our investigation from the paramagnetic (PM) state, where we inhibit magnetic ordering. We find however a CO solution with a charge pattern characterized by alternate charge-poor (P) and charge-rich (R) TMTTF molecules along the stacking direction $a$ as shown in the inset of Fig. \ref{fig1}. The sum of the occupation of two neighboring sites is always three, and the charge disproportionation  $\delta = n_+ - n_-$ ($n_+$ and $n_-$ being the occupancy of the two non-equivalent sites) is such that the R sites approach $n_+=2$ and the P sites approach half-filling. In this light, the present CO state can be interpreted as a Mott-like state\cite{Amaricci}.
In Fig. \ref{fig1} we show the spectral function $A(\omega) = -1/\pi Im G(\omega)$ for the PM-CO phase in the distorted structure with $\xi = 0.01$ which shows an insulating behavior with a charge gap $\Delta$.  For a fixed value of $U$,  ${\Delta}$ increases as a function of $V$ (see inset of the Fig. \ref{fig1}).  For the smallest values of V for which the system is insulating, the theoretical value of $\Delta$ is close to the experimental value\cite{Pashkin}.

We also performed Hartree-Fock calculations in which the Hubbard interaction is treated at mean-field level, and we find that the PM solution has metallic nature, clearly demonstrating the strongly correlated nature of the insulating state and the need to use DMFT to properly account for dynamical correlations. An insulating solution is found in Hartree-Fock only allowing for simultaneous AFM and CO broken symmetries. On the other hand, within DMFT we find a sharp first-order transition from a CO correlated metal to a Mott insulator with charge ordering by increasing V, as obtained for the quarter-filled model in Ref. \cite{Amaricci}. The value of the disproportionation parameter is immediately large ($\delta \simeq 0.6$) as soon as V is sufficient to enter the insulating state. These values are most likely overestimated by the mean-field treatment of the inter-site repulsion in our single-site DMFT approximation, where only  the local dynamical fluctuations are treated exactly. A possible strategy to overcome this limitation would be to use cluster extensions of DMFT. In these methods, however, the mirror symmetry with respect to each site is broken even in the case where the hopping is uniform for even number of sites in the cluster. Therefore we limited to the single-site DMFT which does not introduce a bias in favor of distortion. 

The comparison with experiments confirms the expectations, as all the experimental estimates are smaller than our values. It should however be noticed that nuclear magnetic resonance (NMR) systematically predicts larger values with respect to infrared or Raman spectroscopies. More precisely, NMR gives $\delta = 0.28$ \cite{Nakamura_CO} for X=PF$_6$, 0.33 \cite{Fujiyama_CO} and 0.5 \cite{Zamborszky_CO} for X=AsF$_6$ and 0.5 \cite{Yu} or 0.55 \cite{Iwase_CO} for X=SbF$_6$, while IR gives 0.16, 0.21 and 0.29 respectively \cite{Knoblauch}. However, infrared and Raman estimates are based on several assumptions and they rely on single-molecule estimates of the vibrational frequency, whose accuracy for the present solids is questionable. In this view, the discrepancy between the NMR measurements and our calculations remains significant, but is is acceptable if one takes into account
the limitations of single site DMFT.

\begin{figure}
\includegraphics[width=.6\columnwidth,angle=-90]{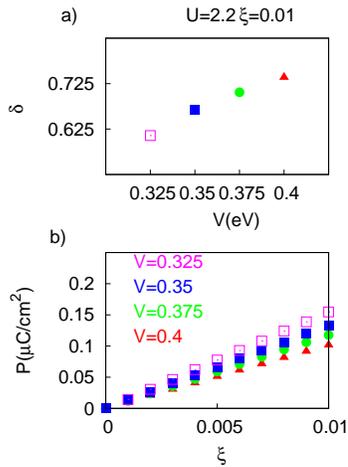}
\vspace{0.25cm}
\caption{(Color online) Charge disproportionation ${\delta}$ (a), (b) and  electronic contribution to the polarization $P$(c), (d) as function of lattice distortions $\xi$ at different values of $V$ and $U$ in the paramagnetic CO state. }
\label{fig2}
\end{figure}

The paramagnetic solution we have found displays simultaneous CO and dimerization which break the inversion symmetry and induces an electric polarization in the $a$ lattice direction \cite{Monceau}. 
In Fig. \ref{fig2}a and \ref{fig2}b we show the computed charge disproportionation $\delta$ and electronic contribution to the polarization, $P$ as a function of $V$ ($P$ is plotted as a function of $\xi$, and different $V$ are compared). Interestingly, $P$ decreases when we increase $V$ thereby increasing $\delta$. This trend is related to the symmetry of the dimerized ferroelectric state which changes from a bond-centered ordering (favored by $\xi$) towards a site-centered ordering (favored by $V$). Note that $\xi=0$, in which site-centered ordering is the ground state has by definition $P=0$.

On the other hand, both $\delta$ and $P$ are not sensitive to $U$ in the range considered in our calculations. It is worth mentioning that however $U$ cooperates with $V$ in opening the insulating gap, confirming the Mott nature of the insulator.

\begin{figure}
\includegraphics[width=.925\columnwidth,angle=0]{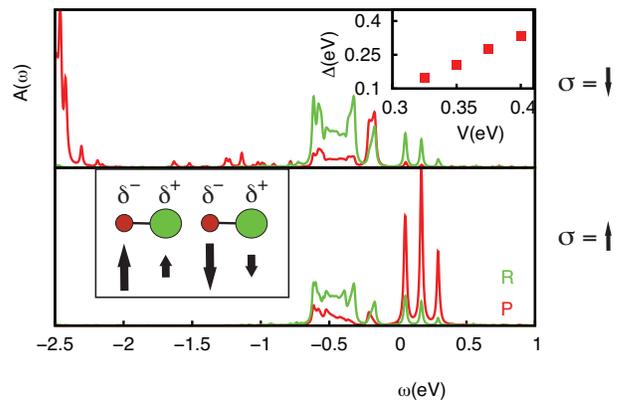}
\caption{(Color online) Spectral density of states calculated in the AFM state for $U=$2.2 and $V=$0.325 within DFT+DMFT. Insets: (Left) Charge and spin arrangements of poor (P) and rich (R) sites in the AFM CO state along the stacking direction $a$ at ${\xi}$=0.1 ; (Right) Charge gap ${\Delta}$ as function of V.}
\label{Fig3}
\end{figure}

\begin{figure}
\vspace{0.25cm}
\includegraphics[width=.6\columnwidth,angle=-90]{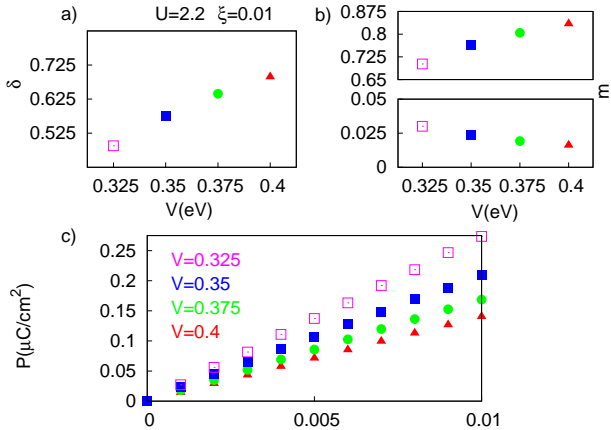}
\caption{(Color online) Charge disproportionation ${\delta}$ (a), (b) on-site magnetization of rich and poor sites $m_{R,P}$ and (c) P electronic contribution to the ferroelectric polarization function of lattice distortions $\xi$ at different values of V in the AFM CO state. }
\label{fig4}
\end{figure}

The charge localization introduced by electronic correlations leads to the formation of local magnetic moments which are typically expected to arrange in some ordered state to minimize the exchange energy. We therefore release the constraint to paramagnetic solutions to study the interplay between ferroelectric and magnetic orders. The spin pattern we obtain is depicted in the inset of Fig.\ref{Fig3}. The charge-rich sites have an occupation which approaches two and a small magnetic moment, while the nearly half-filled charge-poor sites have a large momentum. Within each dimer the two spins are ferromagnetically aligned, while neighboring dimers display antiferromagnetic ordering of the spins. This magnetic structure corresponds to have one effective spin per dimer with an antiferromagnetic inter-dimer coupling and what it has been found in Ref. \cite{Matsunaga} for X=SbF$_6$. As a consequence of the exchange interactions the spin-minority states at the charge-poor sites can be occupied by minority electrons jumping from charge-rich site creating exchange striction of the same kind found in novel inorganic multiferroic materials. When $V$ is reduced the difference between the magnetization at charge-poor and rich sites decreases, thereby enhancing the magnetic striction and the electronic polarization (see Fig. \ref{fig4}c). The spin and charge orders found in our calculations are consistent with the experimental evidence of charge\cite{Monceau} and spin \cite{Dumm,Matsunaga} orderings in TMTTF salts.
In Fig. \ref{Fig3} we show the DMFT spectral function in the AFM-CO phase, which is naturally insulating, with a gap ${\Delta}$ again increasing upon increasing $V$, and very close to its paramagnetic value (see inset Fig. \ref{Fig3}). Fig. \ref{fig4}a) and \ref{fig4}b) show that the disproportionation and the large magnetization $m_P = n_{\uparrow P} - n_{\downarrow P}$ of the charge-poor sites both increase as $V$ increases while the small magnetization $m_R$ of the charge-rich sites decreases. We can therefore conclude that the magnetic ordering is controlled by the charge disproportionation which underlies the ferroelectric behavior. Indeed, the AFM phase favors the formation of a mixed bond-centered/site-centered polar charge ordering. By comparing panel(a) of the Fig. \ref{fig2} and Fig. \ref{fig4}, one can see that the charge disproportionation  is smaller in the AFM phase than in the PM phase and the difference between $\delta$ in the two phases decreases upon increasing $V$ at fixed $U$.
 In the other words, the onset of the  AFM phase shifts the center of negative charge toward the center of bonds with a larger value at smaller $V$ and leads to a larger electric polarization (see Fig. \ref{fig2}b and \ref{fig4}a).  

Similarly to the PM-CO state the electronic polarization decreases with the charge disproportionation ${\delta}$ at the molecular sites and it is linear in the dimerization (see Fig. \ref{fig4}a and \ref{fig4}c).
As pointed out above, magnetism cooperates with the dimerization to go toward a bond-centered/site-centered charge density. 
Therefore, the fact that ferroelectric polarization increases almost by a factor $2$ in the magnetic state is a prediction of the magneto-electric coupling in (TMTTF)$_2$-$X$ crystals.
Our results show that the magnetic state changes following the ferroelectric phase transition which can be varied by applying electric field and inversely, the ferroelectric polarization can be manipulated by applying high magnetic fields. This peculiar magnetoelectric effect in (TMTTF)$_2$-$X$ will be also combined with the magnetoelastic effect that modifies both the superexchange interaction and the molecular bonding.
The enhancement of the electronic contribution to the ferroelectric polarization at the onset of antiferromagnetism appears therefore as a common feature of half-filled correlated insulator \cite{GiovannettiTTFCA,RezaNourafkanPdmft}.

Our calculations predict an electronic polarization $P \simeq$ $0.2$ $\mu$C/cm$^2$ for the actual structure of (TMTTF)$_2$-PF$_6$. The ionic contribution is expected to be smaller because of the centrosymmetric crystal structure.  A quantitative calculation of the electronic polarization, which is still to be determined experimentally, potentially requires the inclusion of more molecular orbitals, non-local quantum fluctuations and a fully microscopic account of the structural distortions. However our conclusions about the magnetoelectric effects are expected to be robust, as they are based on a generic properties of correlated systems with charge and magnetic ordering. We believe that the mechanism we present to be relevant for a wider family of charge-ordered ferroelectric systems as k-(BEDT-TTF)$_2$-Cu[N(CN)$_2$]Cl \cite{Lunkenheimer} and ${\beta}$'-(BEDT-TTF)$_2$ICl$_2$ \cite{Iguchi}. Furthermore, the same magneto-electric coupling we proposed for (TMTTF)$_2$-$X$ may also describe materials in which a low-temperature multiferroic state is replaced at higher temperature by ferroelectric ordering, as observed in other Mn-based multiferroic material \cite{Sakai,GiovannettiSBMO}.

In conclusion, using DFT+DMFT and  a recent method based on Green's function formalism to calculate the electronic contribution to the polarization we investigate the ferroelectric and multiferroic phase of (TMTTF)$_2$-$X$ molecular crystals.
We show that (TMTTF)$_2$-$X$ are strongly correlated insulators developing a charge ordered state which, combined with molecular dimerization, gives a ferroelectric state as experimentally observed. The same correlation effects also drive and antiferromagnetic ordering.
By comparing the values of electronic polarization in the paramagnetic and antiferromagnetic magnetic structure we show that magnetic and ferroelectric orderings are strongly intertwined. These findings establish (TMTTF)$_2$-$X$ to belong to the class of electronic-driven multiferroic materials, and that strong electron-electron correlations is the driving force behind charge-ordering, polarization and magnetic ordering. This kind of Mott state which hosts both charge ordering and antiferromagnetism provides us with a mechanism to overcome the apparent incompatibility of ferroelectric and magnetic ordering.

We are grateful to C. Batista for useful discussions. GG and MC acknowledge financial support by European Research Council under FP7/ERC Starting Independent Research Grant ``SUPERBAD" (Grant Agreement n. 240524). RN and GK are supported by NSF DMR-1308141. Calculations have been performed at CINECA (HPC project lsB06\_SUPMOT).

\end{document}